\documentclass{article}
\usepackage{spconf,amsmath,amssymb,graphicx,booktabs,multirow,siunitx}
\usepackage{microtype}
\usepackage{algorithm}
\usepackage{algpseudocode}
\usepackage{tikz}
\usepackage{hyperref}
\usepackage[x11names]{xcolor}
\definecolor{amber}{RGB}{255,191,0}

\usetikzlibrary{shapes.geometric, arrows.meta, positioning, calc, backgrounds, fit}

\title{Training-Free Instruction TTS Gender Bias Calibration Using Model-Adaptive Steering}

\name{
  Kuan-Yu Chen$^{1,2}$, 
  Yi-Cheng Lin$^{2}$, 
  Jeng-Lin Li$^{1}$, 
  Jian-Jiun Ding$^{2}$\thanks{*Code is available at \url{https://github.com/DanielChen1128/ITTS-Debias}}%
}
\address{
  $^{1}$AI Research Center, Inventec Corporation, Taiwan  \\
  $^{2}$Graduate Institute of Communication Engineering, National Taiwan University, Taiwan \\
}
\begin{document}

\ninept

\maketitle

\begin{abstract}

Instruction text-to-speech (ITTS) systems systematically encode acoustic gender skews from descriptive style prompts, such as occupations or personas, even when demographic attributes are left unspecified. Calibrating the implicit gender distribution of the synthesized voices, across heterogeneous architectures and without model retraining or overriding explicit user prompts, is an open problem. In this work, we propose \textit{model-adaptive steering}, a training-free bias calibration method that steers post-encoder conditioning representations using a group deviation vector paired with a coarse-to-fine strength search on a development set. A deterministic lexical gate bypasses intervention whenever explicit gender keywords are detected, preserving intended prompt semantics. Evaluated on a 12,800-prompt held-out benchmark across four ITTS models (three architectures), the proposed method reduces aggregate calibration error from 11.5--28.1 to 0.8--5.9 percentage points (e.g., shifting Parler-TTS Mini from 78.1\% and PromptTTS++ from 27.3\% female to 50.8--52.4\%). Output rates stay within 0.9 pp across anchor set sizes at fixed operating points, while UTMOS decreases by at most 0.08 and WER degrades by at most 3.3 pp.

\end{abstract}


\section{Introduction}

Instruction text-to-speech (ITTS) enables open-ended natural language control over speaker persona, prosody, and acoustic environments, supporting applications from voice assistants to personalized audiobook synthesis~\cite{guo2023prompttts,yang2024instructtts}. Recent advances have explored ITTS under diverse architectural paradigms, including autoregressive language models~\cite{lyth2024parler}, non-autoregressive acoustic generators~\cite{shimizu2024promptttspp}, and hybrid dual-stream architectures~\cite{zhou2024voxinstruct}. Despite their impressive synthesis quality and stylistic fidelity, these systems exhibit persistent demographic biases. When prompts lack explicit demographic constraints, career, status, or persona attributes, such as ``experienced software engineer'' or ``gentle kindergarten teacher'', can induce skewed acoustic gender profiles~\cite{chen2026binding,kuan2025gender}. Left unmitigated, such implicit defaults risk reinforcing societal stereotypes in downstream deployments.

Although demographic bias has been widely investigated in text processing and speech-related tasks, including automatic speech recognition, speech emotion recognition, and self-supervised representation learning~\cite{koenecke2020racial,tatman2017gender,lin2025emo,lin2025mitigating,tsai2025co,lin2024social}, bias mitigation in generative instruction-guided speech synthesis remains comparatively underexplored. Recently, activation steering has emerged as a training-free mechanism for
controlling the model behavior, first in language models~\cite{turner2023activation,rimsky2024steering,zou2023representation}
and more recently for explicit acoustic attributes such as emotion and
accent~\cite{xie2025emosteer,sun2026activationsteeringaccentadaptation}.
However, existing frameworks steer individual generations toward an explicitly specified attribute, rather than calibrating the aggregate demographic distribution induced by prompts that specify no attribute at all. Naively applying continuous latent shifts also raises two concerns that our design anticipates: overly large offsets may degrade acoustic quality, while
indiscriminate steering may override explicitly specified demographic
attributes. A training-free intervention that reduces demographic disparities
while preserving acoustic fidelity and the user intent thus remains an open
challenge.

In this work, we propose \textit{model-adaptive steering}, a training-free bias calibration method that estimates a separate deviation direction and operating point at each model's own post-encoder conditioning interface. Our approach is based on the assumption that demographic bias manifests as a directional skew in that conditioning space. We estimate an empirical group deviation vector from paired counterfactual anchors to offset the implicit demographic skew and steer the resulting output distribution toward parity. It operates on those representations directly, without parameter-efficient adaptation or model retraining~\cite{gao2022wavprompt,kwon2025parameter}. To mitigate localized demographic saturation under complex semantic combinations, we partition the prompt space into descriptor-type strata~\cite{chen2026binding} and formulate a joint fairness objective across these strata, identifying an operating point $\theta^*$ that avoids stratum saturation through coarse-to-fine grid search.
Finally, a deterministic lexical gate bypasses intervention on gender-specified inputs, activating steering only on prompts without explicit gender keywords. Across a 12,800-prompt held-out test benchmark, our method substantially reduces demographic disparities across prompt strata at a small utility cost.
The main contributions of this work are summarized as follows:
\begin{itemize}
\item \textbf{Training-Free and Model-Adaptive Steering:} A single inference-time intervention, instantiated at each model's own conditioning interface, calibrates gender output rates across three conditioning layouts (token-level, pooled style, dual-stream) without updating generator parameters or adding adapters.

\item \textbf{Multi-Stratum Operating-Point Calibration:} A joint objective over overall and stratum-wise disparities, combined with a saturation filter, prevents the single-gender collapse that aggregate parity alone admits.

\item \textbf{Bias and Utility Evaluation:} Aggregate calibration error drops from 11.5--28.1 to 0.8--5.9 pp across four models, significant against every baseline, at a small utility cost (UTMOS $\le$0.08, WER $\le$3.3 pp).
\end{itemize}

\begin{figure*}[t]
\centering
\resizebox{0.98\textwidth}{!}{%
\begin{tikzpicture}[
    >=Stealth,
    font=\small\sffamily,
    thick,
    % 基礎樣式
    base/.style={rounded corners=3pt, align=center, inner sep=3.5pt},
    % Stage 1 專用樣式
    p1_data/.style={base, draw=cyan!70!black, fill=cyan!8, text width=2.3cm, minimum height=1.2cm},
    p1_proc/.style={base, draw=teal!70!black, fill=teal!8, text width=2.2cm, minimum height=1.2cm},
    p1_calc/.style={base, draw=gray!70!black, fill=gray!8, text width=2.5cm, minimum height=1.2cm},
    p1_search/.style={base, draw=teal!70!black, fill=teal!12, text width=2.6cm, minimum height=1.2cm, line width=1.0pt},
    p1_param/.style={base, draw=teal!85!black, fill=teal!20, text width=2.6cm, minimum height=1.2cm},
    % Stage 2 專用樣式
    p2_data/.style={base, draw=cyan!70!black, fill=cyan!8, text width=1.8cm, minimum height=1.05cm},
    p2_proc/.style={base, draw=teal!70!black, fill=teal!8, text width=1.9cm, minimum height=1.05cm},
    p2_gate/.style={base, draw=amber!85!black, fill=amber!15, rounded corners=5pt, text width=1.6cm, minimum height=1.05cm},
    p2_steer/.style={base, draw=orange!80!black, fill=orange!12, text width=2.5cm, minimum height=1.05cm, line width=1.1pt},
    p2_dec/.style={base, draw=teal!70!black, fill=teal!8, text width=2.0cm, minimum height=1.05cm},
    p2_wav/.style={base, draw=green!60!black, fill=green!10, text width=1.35cm, minimum height=1.05cm},
    p2_eval/.style={base, draw=purple!60!black, fill=purple!5, dashed, text width=2.1cm, minimum height=1.15cm},
    % 背景群組
    group1/.style={draw=teal!50!black, fill=teal!2, dashed, rounded corners=6pt, inner sep=10pt},
    group2/.style={draw=blue!50!black, fill=blue!2, dashed, rounded corners=6pt, inner sep=10pt}
]

% ==================== Stage 1: Offline Calibration ====================
\node[p1_data] (fm_anchors) at (1.1, 3.3) {\textbf{Paired Anchors}\\[1pt]$\mathcal{A}_N = \{(x_j^F, x_j^M, c_j)\}$};
\node[p1_proc] (enc_offline) at (4.7, 3.3) {\textbf{Frozen Text}\\[1pt]Encoder $E$};
\node[p1_calc] (offset_calc) at (8.5, 3.3) {\textbf{Deviation Vector}\\[2pt]$\mathbf{d}$ via Eq.~\eqref{eq:centroid}};
\node[p1_search] (dev_screen) at (12.3, 3.3) {\textbf{Coarse-to-Fine Search}\\[1pt]on $\mathcal{D}_{\text{dev}}$ (\textbf{Sec.~\ref{sec:search}})};
\node[p1_param] (freeze_param) at (16.2, 3.3) {\textbf{Operating Point}\\[2pt]$\theta^*, \mathbf{d}$\\[1pt]\scriptsize{(e.g., $\alpha^*$ or $\alpha_{\text{AR}}^*, \alpha_{\text{NAR}}^*$)}};

% Stage 1 連線
\draw[->] (fm_anchors) -- (enc_offline);
\draw[->] (enc_offline) -- node[above, font=\scriptsize] {$\mathbf{z}(x_j^{M/F})$} (offset_calc);
\draw[->] (offset_calc) -- node[above, font=\scriptsize] {$\mathbf{d}$} (dev_screen);
\draw[->] (dev_screen) -- (freeze_param);

% ==================== Stage 2: Inference Steering ====================
\node[p2_data] (inf_input) at (0.0, 0) {\textbf{Prompt $x$}\\[1pt]+ Text $c$};
\node[p2_proc] (enc_inference) at (2.7, 0) {\textbf{Frozen Text}\\[1pt]Encoder $E$};
\node[p2_gate] (lexical_gate) at (5.3, 0) {\textbf{Lexical Gate}\\[1pt]\scriptsize{$w \in \mathcal{V}_{\text{gender}}$?}};
\node[p2_steer] (intervention) at (8.2, 0) {\textbf{Conditioning Interface}\\[2pt]\footnotesize $\mathbf{h}'_t = \mathbf{h}_t + a_t \cdot \alpha^* \mathbf{d}$};
\node[p2_dec] (decoder) at (11.5, 0) {\textbf{Frozen Decoder}\\[1pt]+ Vocoder};
\node[p2_wav] (wav_out) at (14.0, 0) {\textbf{Speech}\\[2pt]Wave $y$};
% \node[p2_eval] (evaluation) at (16.8, 0) {\textbf{Downstream Eval}\\[1pt]\scriptsize $\bullet$ Parity $|p_F - 0.5|$\\\scriptsize $\bullet$ UTMOS\\\scriptsize $\bullet$ Whisper WER};
\node[p2_eval] (evaluation) at (16.8, 0) {%
    \textbf{Downstream Eval}\\[3pt]
    \scriptsize $\bullet$ Parity $|p_F - 0.5|$\\[4pt]
    \scriptsize $\bullet$ UTMOS\\[2pt]
    \scriptsize $\bullet$ Whisper WER%
};

% Stage 2 連線
\draw[->] (inf_input) -- (enc_inference);
\draw[->] (enc_inference) -- node[above, font=\scriptsize] {$\mathbf{H}$} (lexical_gate);
\draw[->] (lexical_gate) -- node[above, font=\scriptsize] {No} (intervention);
\draw[->] (intervention) -- node[above, font=\scriptsize] {$\mathbf{H}'$} (decoder);
\draw[->] (decoder) -- (wav_out);
\draw[->, dashed] (wav_out) -- (evaluation);

% 參數注入長箭頭
\draw[->, thick, color=orange!85!black] (freeze_param.south) |- ++(0,-0.25) -| node[pos=0.25, above, font=\footnotesize\bfseries] {Inject Operating Point $\theta^*$} (intervention.north);

% Lexical Gate 繞道箭頭
\draw[->, dashed, line width=0.9pt, color=orange!90!black] (lexical_gate.south) -- ++(0,-0.80) -| node[pos=0.25, below, font=\scriptsize] {Yes (Bypass: $g(x) = 0 \Rightarrow \mathbf{h}'_t = \mathbf{h}_t$)} (decoder.south);

% ==================== 背景分組外框 ====================
\begin{scope}[on background layer]
    \node[group1, fit=(fm_anchors) (freeze_param), label={[font=\small\bfseries, text=teal!70!black]above:\strut Stage 1: Offline Group Deviation Estimation \& Operating-Point Calibration}] {};
    \node[group2, fit=(inf_input) (evaluation), label={[font=\small\bfseries, text=blue!70!black]above:\strut Stage 2: Inference-Time Conditional Feature Intervention}] {};
\end{scope}

\end{tikzpicture}%
}

\caption{Two-stage calibration and inference framework. Stage~1 estimates the group deviation vector $\mathbf{d}$ from anchor pairs $\mathcal{A}_N$ and identifies the operating point $\theta^*$ on $\mathcal{D}_{\mathrm{dev}}$ via coarse-to-fine search (Sec.~\ref{sec:search}). Stage~2 shifts post-encoder conditioning representations $\mathbf{H}$ prior to acoustic decoding, where a deterministic lexical gate bypasses steering whenever explicit gender keywords are detected.}
\label{fig:itts_debias_framework}
\end{figure*}
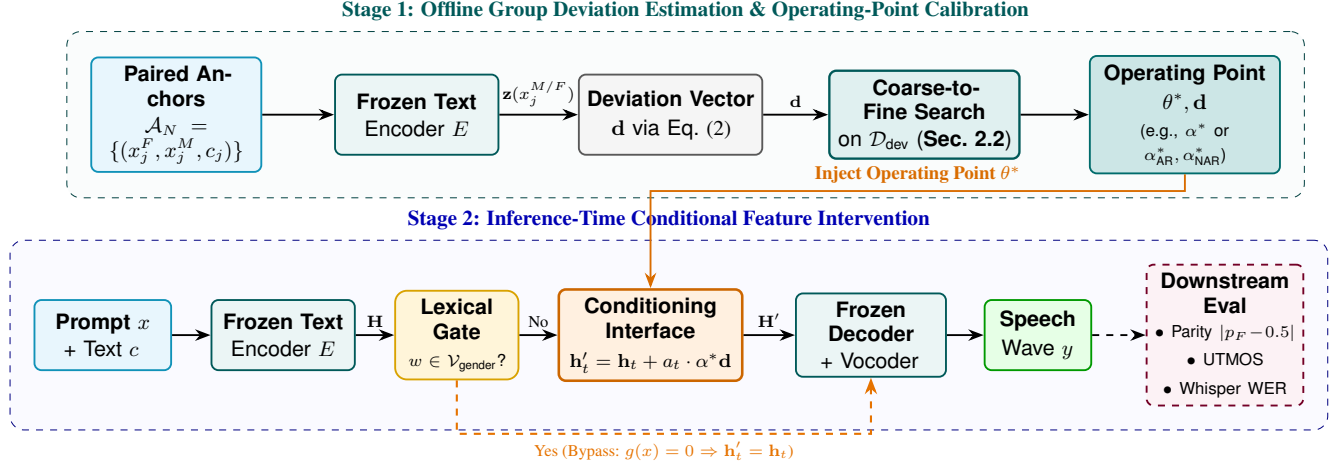

\section{Method}
\vspace{-1mm}

\subsection{Group Deviation Estimation}
\label{sec:deviation_estimation}
Let $M$ denote a frozen ITTS generator that samples speech $y \sim M(x, c)$ from an instruction prompt $x$ and transcript $c$ under fixed seeds. As shown in Fig.~\ref{fig:itts_debias_framework}, the text encoder $E$ maps $(x, c)$ to intermediate conditioning $\mathbf{H}(x, c) = [\mathbf{h}_1, \dots, \mathbf{h}_T] \in \mathbb{R}^{T \times D}$ with attention mask $a_t \in \{0, 1\}$. Parler-TTS and PromptTTS++ encode only $x$; VoxInstruct encodes $x$ and $c$ jointly.

Because different ITTS architectures have distinct representations and native demographic skews, estimation is performed per model. Let $\mathcal{A}_N = \{(x_j^F, x_j^M, c_j)\}_{j=1}^N$ denote $N$ counterfactual anchor pairs sharing identical templates and transcripts but differing solely in explicit gender descriptors (e.g., $x_j^F=$ ``\textit{A woman speaking with a warm tone}'' vs.\ $x_j^M=$ ``\textit{A man speaking with a warm tone}'' paired with identical text $c_j$). Anchors are disjoint from the test set (Sec.~\ref{subsec:dataset}).  Via masked mean pooling over tokens, each sequence is aggregated into
\begin{equation}
    \mathbf{z}(x, c) = \frac{\sum_{t=1}^T a_t \mathbf{h}_t}{\sum_{t=1}^T a_t}.
    \label{eq:pooling}
\end{equation}
From these paired representations, we compute the model-specific \textit{group deviation vector}:
\begin{equation}
    \mathbf{d} = \frac{1}{N}\sum_{j=1}^{N}\left[\mathbf{z}(x_j^M,c_j)-\mathbf{z}(x_j^F,c_j)\right].
    \label{eq:centroid}
\end{equation}
Thus, $\mathbf{d} \in \mathbb{R}^D$ reflects the average female-to-male representation shift, where positive (negative) steering drives features toward male (female) acoustic profiles. Here $\mathbf{d}$ provides an operational steering direction rather than a theoretical disentanglement of gender from correlated semantic attributes. Steering with strength $\alpha\in\mathbb{R}$ shifts every unmasked token:
\begin{equation}
    \mathbf{h}'_t=\mathbf{h}_t+g(x)\,a_t\,\alpha\,\mathbf{d},
    \label{eq:steer}
\end{equation}
where the lexical gate $g(x)=0$ if $x$ contains a term in $\mathcal{V}_{\text{gender}}$ (Sec.~\ref{sec:inference}) and $g(x)=1$ otherwise. $M(x,c;\alpha)$ denotes the generator decoding from $\mathbf{H}'$, thus $\alpha=0$ recovers $M$.

\begin{table*}[t]
\centering
\caption{Held-out evaluation ($12{,}800$) across prompt strata and calibration errors. Error metrics are reported in percentage points (pp), with $p_F$ in \% ($E_{\mathrm{agg}}$ weights strata by size; $E_{\mathrm{macro}}$ weights them equally): $E_{\mathrm{agg}}=|p_F-50|$, $E_{\mathrm{macro}}=\frac{1}{5}\sum_s |p_{F,s}-50|$, and $E_{\mathrm{worst}}=\max_s |p_{F,s}-50|$ ($\downarrow$ lower is better). Speech utility (UTMOS $\uparrow$, WER $\downarrow$) is evaluated on a shared 500-prompt test subset (WER as a fraction). 95\% bootstrap CI half-widths: $\le$0.88 pp (All), $\le$2.00 pp ($E_{\mathrm{macro}}$); 7 utterances $<$0.1\,s are excluded. Test strata hold 100/2,600/3,900/3,100/3,100 prompts, thus status-driven $E_{\mathrm{worst}}$ carries $\pm$9.8 pp. All representation-based methods are fitted on the same 2,500 anchor pairs. \textbf{Bold} denotes the best debiasing result within each model; \underline{underline} denotes the second best (Original serves as the unmitigated baseline).}
\label{tab:full}
\footnotesize
\renewcommand{\arraystretch}{1.08}
\setlength{\tabcolsep}{3.0pt}
\begin{tabular}{@{}ll *{6}{S[table-format=2.2]} *{3}{S[table-format=2.2]} S[table-format=1.4] S[table-format=1.4]@{}}
\toprule
& & \multicolumn{6}{c}{Female Classification Rate (\%)} 
  & \multicolumn{3}{c}{Calibration Error (pp) $\downarrow$} 
  & \multicolumn{2}{c}{Utility} \\
\cmidrule(lr){3-8} \cmidrule(lr){9-11} \cmidrule(lr){12-13}
Model & Setting 
& {All} & {Status} & {Career} & {Persona} & {2-axis} & {3-axis}
& {$E_{\mathrm{agg}}$} & {$E_{\mathrm{macro}}$} & {$E_{\mathrm{worst}}$}
& {UTMOS $\uparrow$} & {WER $\downarrow$} \\
\midrule

\multirow{4}{*}{Parler-Mini}
& Original
& 78.11 & 82.00 & 79.42 & 76.92 & 78.77 & 77.71
& 28.11 & 28.97 & 32.00
& 3.8346 & 0.1826 \\

& Neutral
& 86.07 & {\underline{76.00}} & 85.08 & 88.13 & 85.26 & 85.45
& 36.07 & 33.98 & 38.13
& 3.9021 & 0.1314 \\

& LEACE$_{\text{pb}}$ ($\beta=1.00$)
& {\underline{74.84}} & 77.00 & {\underline{83.08}} & {\underline{67.33}} & {\underline{78.03}} & {\underline{74.13}}
& {\underline{24.84}} & {\underline{25.91}} & {\underline{33.08}}
& 3.8512 & 0.1419 \\

& Ours ($\alpha=2.5$)
& {\textbf{52.43}} & {\textbf{51.00}} & {\textbf{56.62}} & {\textbf{46.79}} & {\textbf{54.94}} & {\textbf{53.55}}
& {\textbf{2.43}} & {\textbf{3.86}} & {\textbf{6.62}}
& 3.8059 & 0.1413 \\

\addlinespace[3.5pt]

\multirow{4}{*}{Parler-Large}
& Original
& 74.83 & 53.00 & 88.46 & 85.69 & 63.99 & 61.29
& 24.83 & 20.49 & 38.46
& 3.5362 & 0.2487 \\

& Neutral
& 79.51 & {\textbf{62.00}} & 92.92 & 84.58 & 72.19 & 69.77
& 29.51 & 26.29 & 42.92
& 3.2827 & 0.3224 \\

& LEACE$_{\text{pb}}$ ($\beta=0.50$)
& {\underline{75.39}} & 68.00 & {\underline{90.00}} & {\underline{83.68}} & {\underline{65.87}} & {\textbf{62.45}}
& {\underline{25.39}} & {\underline{24.00}} & {\underline{40.00}}
& 3.5427 & 0.2403 \\

& Ours ($\alpha=2.0$)
& {\textbf{51.31}} & {\underline{36.00}} & {\textbf{71.81}} & {\textbf{61.05}} & {\textbf{39.84}} & {\underline{33.84}}
& {\textbf{1.31}} & {\textbf{14.64}} & {\textbf{21.81}}
& 3.4601 & 0.2813 \\

\addlinespace[3.5pt]

\multirow{4}{*}{PromptTTS++}
& Original
& 27.28 & 0.00 & 53.15 & 18.95 & 24.10 & 20.13
& 22.72 & 28.00 & 50.00
& 3.6751 & 0.0064 \\

& Neutral
& 38.92 & 4.00 & 77.77 & {\underline{39.54}} & 26.81 & 18.81
& 11.08 & 27.72 & 46.00
& 3.6328 & 0.0039 \\

& LEACE$_{\text{pb}}$ ($\beta=0.50$)
& {\underline{58.20}} & {\underline{12.00}} & {\underline{63.12}} & 69.33 & {\underline{51.84}} & {\textbf{47.90}}
& {\underline{8.20}} & {\underline{14.88}} & {\underline{38.00}}
& 3.7247 & 0.0077 \\

& Ours ($\alpha=-0.25$)
& {\textbf{50.80}} & {\textbf{41.00}} & {\textbf{56.88}} & {\textbf{51.15}} & {\textbf{48.94}} & {\underline{47.45}}
& {\textbf{0.80}} & {\textbf{4.13}} & {\textbf{9.00}}
& 3.5999 & 0.0130 \\

\addlinespace[3.5pt]

\multirow{4}{*}{VoxInstruct}
& Original
& 61.47 & 58.00 & 61.37 & 65.18 & 62.23 & 56.26
& 11.47 & 10.61 & 15.18
& 2.2949 & 0.3178 \\

& Neutral
& {\underline{56.81}} & {\textbf{45.00}} & 67.85 & 65.67 & {\textbf{50.52}} & {\underline{43.10}}
& {\underline{6.81}} & {\underline{9.19}} & 17.85
& 2.3671 & 0.5253 \\

& LEACE$_{\text{pb}}$ ($\beta=0.75$)
& 61.87 & 64.00 & {\underline{60.88}} & {\underline{64.02}} & 63.32 & 58.48
& 11.87 & 12.14 & {\underline{14.02}}
& 2.2825 & 0.2778 \\

& Ours (AR/NAR: 4, 3)
& {\textbf{55.86}} & {\underline{56.00}} & {\textbf{55.31}} & {\textbf{60.92}} & {\underline{55.23}} & {\textbf{50.58}}
& {\textbf{5.86}} & {\textbf{5.61}} & {\textbf{10.92}}
& 2.2925 & 0.2952 \\

\bottomrule
\end{tabular}
\end{table*}

\subsection{Coarse-to-Fine Steering Strength Search}
\label{sec:search}

Steering strength is calibrated exclusively on the 500-item development set $\mathcal{D}_{\text{dev}}$, holding the deviation vector $\mathbf{d}$ ($N=2{,}500$) frozen. We group prompts into five \textit{strata} $\mathcal{S}$ by descriptor type: status, career, persona, and two-/three-axis prompts combining two or three of these; $\mathcal{D}_{\text{dev}}$ holds 100 items $\mathcal{D}_s$ per stratum $s$. Let $y_i(\theta)\sim M(x_i,c_i;\theta)$ and $G(y)\in\{0,1\}$ denote the binary acoustic-gender proxy ($1=\text{female}$, Eq.~\eqref{eq:gender_decision}). The steering strength is $\theta=\alpha$ for single-stream models and $\theta=(\alpha_{\mathrm{AR}},\alpha_{\mathrm{NAR}})$ for dual-stream VoxInstruct. We monitor overall and stratum-wise female classification rates:
\begin{equation}
\begin{split}
p_F(\theta) &= \frac{1}{|\mathcal{D}_{\text{dev}}|} \sum_{i\in\mathcal{D}_{\text{dev}}} \mathbb{I}[G(y_i(\theta))=1], \\
p_{F,s}(\theta) &= \frac{1}{|\mathcal{D}_s|} \sum_{i\in\mathcal{D}_s} \mathbb{I}[G(y_i(\theta))=1], \quad \forall s \in \mathcal{S}.
\end{split}
\label{eq:rates}
\end{equation}

To avoid single-gender collapse, candidates causing stratum saturation ($p_{F,s}(\theta)\in\{0,1\}$) are discarded. Remaining configurations are evaluated by minimizing the joint parity objective:
\begin{equation}
J(\theta)=|p_F(\theta)-0.5|+\frac{1}{|\mathcal{S}|}\sum_{s\in\mathcal{S}}|p_{F,s}(\theta)-0.5|.
\label{eq:selection}
\end{equation}
The first term alone would let opposite stratum errors cancel; the second penalizes them individually. Rates are fractions in \eqref{eq:rates}--\eqref{eq:selection} and percentages in Tables~\ref{tab:full}--\ref{tab:pairs}.

As discrete proxy classifications render $J(\theta)$ non-differentiable, candidate configurations are optimized via coarse-to-fine grid search. We first evaluate a broad coarse grid $\Theta_{\text{coarse}}$ to bound the parity region, followed by an optional evaluation of finer candidates $\Theta_{\text{refine}}$ within that region.

For the three single-stream models, we evaluate coarse candidates $\Theta_{\text{coarse}}=\{0,\allowbreak\pm1.0,\allowbreak\pm1.5,\allowbreak\pm2.0,\allowbreak\pm2.5\}$. For PromptTTS++, coarse screening is followed by the refinement set $\Theta_{\text{refine}}=\{-0.375,\allowbreak-0.300,\allowbreak-0.250,\allowbreak-0.200,\allowbreak-0.125\}$. For VoxInstruct, we search a coarse $4\times4$ grid with $(\alpha_{\mathrm{AR}},\alpha_{\mathrm{NAR}})\in\{0,2,4,6\}\times\{0,1,2,3\}$. Ties in $J(\theta)$ are broken by selecting the parameter with the lowest worst-case stratum deviation $E_{\mathrm{worst}}(\theta)=\max_s |p_{F,s}(\theta)-0.5|$, followed by the minimal parameter norm $\|\theta\|_2$. The result is frozen as the calibrated \textit{operating point} $\theta^*$ and evaluated unchanged on the held-out test data.

% The chosen strength remains fixed during sensitivity checks. All primary evaluations use the 2,500-pair direction; subsets of 500 and 1,500 pairs are examined solely to assess sample efficiency.

\subsection{Inference-Time Conditional Feature Intervention}
\label{sec:inference}

During inference, a deterministic lexical gate first scans prompt $x$ via case-insensitive word-boundary matching against gender lexicon $\mathcal{V}_{\text{gender}}$ (\textit{male}, \textit{female}, \textit{man}, \textit{woman}, \textit{boy}, \textit{girl}). Detecting any keyword $w\in\mathcal{V}_{\text{gender}}$ sets $g(x)=0$ and bypasses intervention, preserving explicit user intent. For unspecified prompts, \eqref{eq:steer} is applied with the operating point $\theta^*$ at each generator's own conditioning interface:

\noindent\textbf{Token-Level (Parler-TTS):} \eqref{eq:steer} with $\alpha=\alpha^*$, applied prior to acoustic cross-attention.

\noindent\textbf{Global Style (PromptTTS++):} \eqref{eq:steer} with $\alpha=\alpha^*$ and $T=1$, applied to the pooled style embedding.

\noindent\textbf{Dual-Stream (VoxInstruct):} Applied independently to the autoregressive (AR) and non-autoregressive (NAR) conditioning streams, using separate deviation vectors $(\mathbf{d}_{\mathrm{AR}},\mathbf{d}_{\mathrm{NAR}})$ (each via \eqref{eq:centroid}) and operating point $(\alpha_{\mathrm{AR}}^*,\alpha_{\mathrm{NAR}}^*)$:
\begin{equation}
\mathbf{h}'_{k,t}=\mathbf{h}_{k,t}+g(x)\,a_{k,t}\,\alpha_k^*\mathbf{d}_k,\qquad k\in\{\mathrm{AR},\mathrm{NAR}\}.
\end{equation}
In all settings, generator parameters remain frozen; we modify only the conditioning features at inference.

\subsection{Baselines and Evaluation Metrics}
\label{sec:baselines_metrics}

\noindent\textbf{Parameter-Update-Free Baselines.}
Under identical data splits, we benchmark against two training-free controls: (i) a \textit{prompt-level} baseline (Neutral) appending ``\textit{The speaker's gender is unspecified; do not infer it from occupation or persona}'', and (ii) a \textit{representation-level} baseline, LEACE$_{\text{pb}}$, applying LEACE~\cite{belrose2023leace} to pooled features, $\mathbf{z}'=(1-\beta)\mathbf{z}+\beta T_{\mathrm{LEACE}}(\mathbf{z})$, with $\mathbf{z}'-\mathbf{z}$ broadcast to tokens as in \eqref{eq:steer}; LEACE is fitted on pooled anchor features and gender labels, and $\beta\in[0,1]$ is tuned via \eqref{eq:selection}.

\noindent\textbf{Acoustic Proxy Classifier.}
Acoustic gender is assessed via a pretrained wav2vec~2.0 classifier~\cite{baevski2020wav2vec,burkhardt2023agegender} yielding logits $(q_F,q_M,q_C)$. Following established audit protocols~\cite{chen2026binding,kuan2025gender}, we derive binary proxy predictions while excluding child logit $q_C$:
\begin{equation}
G(y)=\mathbb{I}\left(q_F(y)\ge q_M(y)\right).
\label{eq:gender_decision}
\end{equation}
Utterances whose top class is child are thus still labeled F or M. $G(y)$ serves strictly as an automated acoustic proxy for synthetic output distributions rather than ground-truth speaker identity.

\noindent\textbf{Speech Utility Measures.}
To assess potential utility trade-offs, acoustic naturalness is measured via UTMOS22~\cite{saeki2022utmos}, and transcription intelligibility via Word Error Rate (WER) using Whisper-large-v3~\cite{radford2023whisper}.

\noindent\textbf{Statistical and Diagnostic Analyses.}
95\% confidence intervals (CIs) come from a stratified bootstrap~\cite{hesterberg2011bootstrap} over test prompts (10{,}000 resamples); setting differences use a paired bootstrap, and utility changes a paired Wilcoxon signed-rank test. For each descriptor with $\ge$10 prompts (multi-axis prompts count toward each component), we compute its female rate and report Spearman's $\rho$ between Original and steered rates, its noise ceiling from binomial redraws, and the standard deviation (s.d.) of rates. We also report the fraction of outputs with binary posterior $P_F=\sigma(q_F-q_M)\in[0.4,0.6]$.
\section{Experiments}

\begin{table}[t]
\centering
\caption{Female classification rates (\%) on the 500-item development set. \textbf{Bold} with an asterisk (*) indicates the selected operating point $\theta^*$.}
\label{tab:alpha_female_screen}
\footnotesize
\renewcommand{\arraystretch}{1.12}

% ==================== Panel (a): Coarse Grid ====================
\textbf{(a) Coarse Grid Search ($\alpha \in [-2.5, 2.5]$)} \\[0.25em]

\setlength{\tabcolsep}{2.0pt}
\begin{tabular}{@{}l *{9}{c}@{}}
\toprule
Model & {$-2.5$} & {$-2.0$} & {$-1.5$} & {$-1.0$} & {$0.0$} & {$1.0$} & {$1.5$} & {$2.0$} & {$2.5$} \\
\midrule
Parler-Mini  & 96.8 & 96.6 & 94.4 & 90.6 & 78.6 & 66.2 & 62.4 & 55.0 & \textbf{53.2}$^*$ \\
Parler-Large & 89.0 & 86.8 & 84.4 & 83.4 & 73.4 & 65.8 & 59.2 & \textbf{51.0}$^*$ & 44.4 \\
PromptTTS++  & 99.6 & 100.0 & 100.0 & 100.0 & 26.2 & 0.0 & 0.0 & 0.0 & 0.0 \\
\bottomrule
\end{tabular}

\vspace{0.75em}

% ==================== Panel (b): PromptTTS++ Local ====================
\textbf{(b) PromptTTS++ Local Refinement Grid} \\[0.25em]
\setlength{\tabcolsep}{6.0pt}
\begin{tabular}{@{}lccccc@{}}
\toprule
$\alpha$ & $-0.375$ & $-0.300$ & $\mathbf{-0.250^*}$ & $-0.200$ & $-0.125$ \\
\midrule
Female (\%) & 67.6 & 56.6 & \textbf{48.6} & 41.0 & 33.8 \\
\bottomrule
\end{tabular}

\vspace{0.75em}

% ==================== Panel (c): VoxInstruct 2D Grid ====================
\textbf{(c) VoxInstruct Dual-Stream Grid ($\alpha_{\mathrm{AR}} \times \alpha_{\mathrm{NAR}}$)} \\[0.25em]
\setlength{\tabcolsep}{8.5pt}
\begin{tabular}{@{}lcccc@{}}
\toprule
$\alpha_{\mathrm{AR}} \;\backslash\; \alpha_{\mathrm{NAR}}$ & $\mathbf{0}$ & $\mathbf{1}$ & $\mathbf{2}$ & $\mathbf{3}$ \\
\midrule
$\mathbf{0}$ & 63.4 & 63.4 & 63.2 & 63.6 \\
$\mathbf{2}$ & 62.6 & 63.0 & 62.4 & 62.4 \\
$\mathbf{4}$ & 60.2 & 59.6 & 59.8 & \textbf{59.4}$^*$ \\
$\mathbf{6}$ & 61.4 & 60.8 & 60.8 & 60.8 \\
\bottomrule
\end{tabular}
\end{table}

\begin{table}[t]
\centering
\caption{Sensitivity of female classification rates (\%) to anchor pair count $N$ on the held-out test set ($12{,}800$). Calibrated operating points $\theta^*$ remain fixed. $\Delta$ denotes the maximum percentage point variation across pair counts.}
\label{tab:pairs}
\small
\renewcommand{\arraystretch}{1.12}
\setlength{\tabcolsep}{8.0pt}
\begin{tabular}{@{}lcccc@{}}
\toprule
& \multicolumn{3}{c}{Anchor Pairs ($N$)} & \\
\cmidrule(lr){2-4}
Model & {500} & {1{,}500} & {2{,}500} & {$\Delta$ (pp)} \\
\midrule
Parler-Mini  & 52.60 & 52.82 & 52.43 & 0.39 \\
Parler-Large & 52.00 & 52.19 & 51.31 & 0.88 \\
PromptTTS++  & 51.20 & 50.50 & 50.80 & 0.70 \\
VoxInstruct  & 56.30 & 56.06 & 55.86 & 0.44 \\
\bottomrule
\end{tabular}
\end{table}

\subsection{Experimental Setup and Evaluation Protocol}
\label{subsec:dataset}

Our benchmark comprises 13,300 instruction prompts partitioned into a 500-item development set ($\mathcal{D}_{\text{dev}}$) and a 12,800-item test set. The development set is balanced across the five strata (Sec.~\ref{sec:search}; 100 prompts each). The held-out test partition contains the remaining 100 status, 2,600 career, 3,900 persona, 3,100 two-axis, and 3,100 three-axis prompts. This division enforces strict prompt-ID disjointness between calibration and evaluation, although lexical descriptors and templates originate from the shared benchmark inventory. Across all evaluated systems, generation seeds, target transcripts, and batching configurations are held constant to ensure identical synthesis conditions across interventions.

Both model-adaptive steering and LEACE$_{\text{pb}}$ are fitted on the same 2,500 paired anchors: 500 derive from $\mathcal{D}_{\text{dev}}$ prompts, extended with self-constructed pairs to 1,500 and 2,500. Calibration parameters ($\theta^*$ and $\beta^*$) are optimized exclusively on $\mathcal{D}_{\text{dev}}$ and frozen prior to generating and evaluating the 12,800 held-out test prompts, preventing data leakage. Acoustic gender proxy classifications follow the evaluation setup in Sec.~\ref{sec:baselines_metrics}, and speech utility metrics (UTMOS and WER) are evaluated as absolute means on the same 500 test prompts for all settings.

\subsection{Operating-Point Calibration}
As shown in Table~\ref{tab:alpha_female_screen}, unmitigated models ($\alpha=0$) display pronounced, architecture-dependent baseline skews on the development set: Parler-Mini and Parler-Large heavily favor female classifications, whereas PromptTTS++ skews toward male. Calibration via \eqref{eq:selection} counteracts these polarities while avoiding stratum saturation. VoxInstruct starts at 63.4\%. The selected points bring Parler-Mini, Parler-Large, and PromptTTS++ to 53.2\%, 51.0\%, and 48.6\%, and VoxInstruct to 59.4\% at $(\alpha_{\mathrm{AR}}^*, \alpha_{\mathrm{NAR}}^*)=(4,3)$; Parler-Mini and $\alpha_{\mathrm{NAR}}^*$ sit at their grid edges. In our runs the stratum term never altered the selected point; its operative role was discarding saturated candidates. LEACE$_{\text{pb}}$ weights $\beta^*$ (Table~\ref{tab:full}) are selected with the same objective.

\subsection{Held-Out Evaluation and Robustness Analysis}
\noindent\textbf{Debiasing Efficacy Across Architectures.}
As reported in Table~\ref{tab:full}, model-adaptive steering substantially reduces calibration errors across all four systems on the held-out test partition. Aggregate error ($E_{\mathrm{agg}}$) drops from 11.47--28.11 to 0.80--5.86 percentage points (pp), while consistently attaining the lowest point-estimate errors across macro ($E_{\mathrm{macro}}$) and worst-stratum ($E_{\mathrm{worst}}$) metrics. Paired bootstrap CIs of the $E_{\mathrm{macro}}$ reduction exclude zero for all four models, both against Original and against the best baseline. In contrast, appending neutrality instructions exacerbates female skew in Parler models. While LEACE$_{\text{pb}}$ lowers aggregate error in PromptTTS++ ($E_{\mathrm{agg}} = 8.20$ pp), it under-corrects status (12.00\% female) while shifting persona to 69.33\%.

\noindent\textbf{Anchor-Count Sensitivity.}
Table~\ref{tab:pairs} evaluates sensitivity to anchor subset size ($N \in \{500, 1{,}500, 2{,}500\}$) at fixed operating points. Downstream female classification rates vary within narrow margins across the tested nested anchor subsets, with the maximum shift across counts ($\Delta$) bounded within 0.88 pp across all systems. Though based on one nested sequence per model, this suggests direction estimation is robust to anchor reduction.

\subsection{Stratum-Level Consistency and Speech Utility}
\noindent\textbf{Stratum-Level Parity Breakdown.}
Aggregate balance does not ensure uniform calibration across strata. Our method keeps $E_{\mathrm{worst}}$ within 6.62--10.92 pp except for Parler-Large. Parler-Large retains $E_{\mathrm{worst}} = 21.81$ pp in career, consistent with its large baseline spread (38.46 pp), which a single global offset mitigates but cannot eliminate.

\noindent\textbf{Descriptor-Level Analysis.}
Spearman's $\rho$ between Original and steered descriptor rates is 0.91--0.92 for Parler (binomial noise ceiling 0.93--0.95), 0.69 for PromptTTS++ (0.98), and 0.82 for VoxInstruct. The mean descriptor rate moves toward 50\% in every model, yet the s.d.\ across descriptors widens for both Parler models (13.9$\to$17.8, 18.1$\to$25.5 pp), narrows for PromptTTS++ (32.6$\to$10.6), and is flat for VoxInstruct: a global offset can widen descriptor spread.

\noindent\textbf{Classifier Posterior.}
The fraction of ambiguous outputs ($P_F\in[0.4,0.6]$) is 2.7\% under Original and 3.4\% under steering (mean over models). Parity thus reflects a mix of clearly female and clearly male voices rather than ambiguous outputs.

\noindent\textbf{Speech Utility Evaluation.}
On this 500-prompt subset (Table~\ref{tab:full}), objective speech utility shifts remain modest. Relative to Original, UTMOS decreases by at most 0.076, and WER ranges from $-4.13$ pp (Parler-Mini) to $+3.26$ pp (Parler-Large). LEACE$_{\text{pb}}$ attains higher UTMOS on three of four models and lower WER on three of four; our method trades this small utility cost for substantially lower calibration error. Paired Wilcoxon tests find significant ($p<0.05$) UTMOS changes for 1 of 4 models and WER changes for 2 of 4 (median $\Delta$WER $=0$); perceptual evaluation is left for future work.
\section{Conclusion}
We presented model-adaptive steering, a training-free method to calibrate acoustic gender output rates in frozen ITTS systems via post-encoder conditioning offsets. Across 12,800 held-out prompts, it reduces calibration errors across three conditioning layouts at a cost of at most 0.08 UTMOS, attaining lower point-estimate errors than both baselines. At fixed operating points, output rates stay within 0.9 pp down to 500 anchor pairs. While Parler-Large and VoxInstruct retain residual stratum disparities, our findings show that classifier-assigned demographic outputs can be modulated via simple offsets in conditioning space.

% ----------------------------------------------------
% 第 5 頁：Ethics Statement 與 References 共享
% ----------------------------------------------------
\clearpage

\section*{Compliance with Ethical Standards}
We highlight several ethical considerations and methodological boundaries in this work. First, our evaluation operationalizes gender as a binary construct ($F/M$) to interface with available pretrained acoustic models and annotations; this engineering simplification does not reflect the full, non-binary spectrum of human gender identity. Second, our study examines perceived acoustic gender in isolation, omitting intersecting demographic dimensions such as age, ethnicity, and regional dialect. Third, all benchmarks, lexicons, and models are strictly restricted to English, warranting caution when generalizing our conclusions to languages with explicit grammatical gender or distinct cultural associations. Finally, the automated classifier $G(y)$ functions strictly as an acoustic proxy to characterize statistical skews across synthetic output distributions, and must not be interpreted as assessing or validating true personal identities.

\footnotesize
\bibliographystyle{IEEEbib}
\bibliography{refs}

\end{document}